\documentclass[10pt]{article}

\usepackage{bbm}
\usepackage[final]{graphics}
\usepackage{amsmath}
\usepackage{amsfonts,amsbsy}
\usepackage{amssymb}
\def\empile#1\over#2{\mathrel{\mathop{\kern 0pt#1}\limits_{#2}}}

\newcommand{\slv}{\raise.15ex\hbox{$/$}\kern-.53em\hbox{$v$}}
\newcommand{\slF}{\raise.15ex\hbox{$/$}\kern-.53em\hbox{$F$}}
\newcommand{\slL}{\raise.15ex\hbox{$/$}\kern-.53em\hbox{$L$}}
\newcommand{\slP}{\raise.15ex\hbox{$/$}\kern-.53em\hbox{$P$}}
\newcommand{\slp}{\raise.15ex\hbox{$/$}\kern-.53em\hbox{$p$}}
\newcommand{\slq}{\raise.15ex\hbox{$/$}\kern-.53em\hbox{$q$}}
\newcommand{\slR}{\raise.15ex\hbox{$/$}\kern-.53em\hbox{$R$}}
\newcommand{\slQ}{\raise.15ex\hbox{$/$}\kern-.53em\hbox{$Q$}}
\newcommand{\slK}{\raise.15ex\hbox{$/$}\kern-.53em\hbox{$K$}}
\newcommand{\slk}{\raise.15ex\hbox{$/$}\kern-.53em\hbox{$k$}}
\newcommand{\slD}{\raise.15ex\hbox{$/$}\kern-.53em\hbox{$D$}}
\newcommand{\slC}{\raise.15ex\hbox{$/$}\kern-.53em\hbox{$C$}}
\newcommand{\slA}{\raise.15ex\hbox{$/$}\kern-.53em\hbox{$A$}}
\newcommand{\slSigma}{\raise.15ex\hbox{$/$}\kern-.53em\hbox{$\Sigma$}}
\newcommand{\slpartial}{\raise.15ex\hbox{$/$}\kern-.53em\hbox{$\partial$}}
\newcommand{\slcalP}{\raise.15ex\hbox{$/$}\kern-.63em\hbox{$\cal P$}}

\def\p{{\boldsymbol p}}
\def\q{{\boldsymbol q}}

\def\k{{\boldsymbol k}}

\def\x{{\boldsymbol x}}

\title{\bf Particle production and AGK relations\\ in the Color Glass Condensate picture}

\author{Fran\c{c}ois Gelis${}^{1}$\thanks{francois.gelis@cea.fr},
     \ \ Raju Venugopalan${}^{2}$\thanks{raju@bnl.gov}}

\begin{document}

\maketitle

\begin{enumerate}
\item {Service de Physique Th\'eorique (URA 2306 du CNRS)\\
    CEA/DSM/Saclay, 91191, Gif-sur-Yvette Cedex, France}%
\item {Nuclear Theory, Physics Department\\
    Brookhaven National Laboratory, Upton, NY 11973, USA}%
\end{enumerate}

\begin{abstract}In this talk, we discuss some general properties of
  particle production in a field theory coupled to strong time
  dependent sources, and techniques to compute the spectrum of the
  produced particles in such theories. We also discuss the application
  of these results to the description of hadron or heavy ion
  collisions in the Color Glass Condensate framework.
\end{abstract}

\section{Introduction}
\label{sec:intro}
At high energy, all the internal timescales of a hadron are time
dilated. Therefore, more and more soft fluctuations -- carrying a
smaller and smaller fraction $x$ of the hadron momentum -- become
long-lived and become relevant in interactions with another hadron. On
the contrary, on the timescales relevant for such an interaction
process, the large $x$ partons can be seen as completely frozen
degrees of freedom, whose only role is to act as sources that radiate
more small $x$ gluons. Moreover, the small $x$ modes will eventually
have an occupation number larger than unity, and undergo
recombinations -- a process known as {\sl saturation} \cite{GriboLR1}.

In the Color Glass Condensate (CGC) framework \cite{MV,CGC}, one thus
divides the degrees of freedom of a hadron in static color sources --
described by a density denoted $\rho$ -- that represent the large $x$
partons, and dynamical gauge fields that represent the small $x$
partons. The CGC can thus be seen as an effective theory of gauge
fields coupled to external sources. The details of this separation of
degrees of freedom can change with the separation scale, but this
should not affect physical quantities. This leads to a renormalization
group equation -- known as the JIMWLK equation \cite{CGC} -- that
governs the evolution with $x$ of the distribution $W[\rho]$ of hard
color sources.

In these proceedings, we consider the collision at high energy of two
hadrons (or heavy ions) described in the CGC framework. We assume that
the distributions of hard color sources that describe the two
projectiles are known, and we address the question of calculating
physical observables in given configurations of the two sources.
Moreover, we will consider only the regime where the two projectiles
are saturated, in which the two sources $\rho_{1,2}$ are strong --
both of order $g^{-2}$.

As stated before, one must consider an effective theory described by
the following Lagrangian,
\begin{equation}
{\cal L}\equiv -\frac{1}{2}\,{\rm tr}\,F_{\mu\nu}F^{\mu\nu}
+
A_\mu \left(J^\mu_1 + J_2^\mu\right)\; ,
\end{equation}
with currents given at lowest order in the sources by
\begin{equation}
J_1^\mu =g\delta^{\mu+}\delta(x^-)\rho_1(\x_\perp)\quad,\qquad
J_2^\mu =g\delta^{\mu-}\delta(x^+)\rho_2(\x_\perp)\; .
\label{eq:J}
\end{equation}
These currents must be covariantly conserved --
$\big[D_\mu,J_{1,2}^\mu\big]=0$ -- which leads to a feedback of the
gauge field on the currents. Therefore, in general, the
eqs.~(\ref{eq:J}) get modified by corrections of higher order in
$\rho_{1,2}$. An important observation is that the strength of the
sources lead to non-perturbative effects, in the sense that an
infinite set of diagrams must be summed in order to calculate a
quantity at a fixed order in $g$. However, the large strength of the
sources has also a valuable consequence: as we shall see later, the
leading order is dominated by tree diagrams only, and it can be
studied by classical methods.

\section{AGK cancellations}
\label{sec:agk}
Following \cite{GelisV}, let us first consider the theory of a real
scalar field coupled to strong sources. Most of the {\sl structural}
properties we want to discuss can indeed already be studied in this
simpler framework. Some general results -- that are to a large extent
independent of the details of the theory under consideration -- can be
obtained by considering the generating function for the probabilities
$P_n$ of producing a given number $n$ of particles~:
\begin{equation}
{\cal F}(z)\equiv\sum_{n=0}^{\infty} P_n\;z^n\; .
\label{eq:F-def}
\end{equation}
In \cite{GelisV}, we have proven that, if we denote $b_r/g^2$ the sum
of all the cut {\sl connected} vacuum-vacuum diagrams\footnote{The
  explicit factor $1/g^2$ represents the natural order of these cut
  diagrams when the sources are strong.} with exactly $r$ cut lines,
the logarithm of ${\cal F}(z)$ reads
  \begin{equation}
   \ln {\cal F}(z)=\frac{1}{g^2}\sum_{r=1}^{\infty}b_r\;(z^r-1)\; .
  \label{eq:F1}
\end{equation} 
From eqs.~(\ref{eq:F-def}) and (\ref{eq:F1}), one gets the following
expression\footnote{A model for the coefficients $b_r$ has recently
been proposed in \cite{Dremin}.} for the probability $P_n$~:
\begin{equation}
P_n=e^{-\sum_r b_r/g^2}\; 
\sum_{p=0}^n \frac{1}{p!}\;
\sum_{r_1+\cdots+r_p=n}
\frac{b_{r_1}\cdots b_{r_p}}{g^{2p}}\; .
\end{equation}
In this formula, the index $p$ is the number of connected cut
subdiagrams from which the $n$ particles are produced: $r_1$ particles
are produced by the first of these $p$ subdiagrams, $r_2$ by the
second one, etc... If we further expand the prefactor $\exp(-\sum_r
b_r/g^2)$, we obtain~:
\begin{equation}
P_n=
\sum_{q=0}^\infty\frac{(-1)^q}{q!}\Big(\sum_r \frac{b_r}{g^2}\Big)^q
\sum_{p=0}^n \frac{1}{p!}\;
\sum_{r_1+\cdots+r_p=n}
\frac{b_{r_1}\cdots b_{r_p}}{g^{2p}}\; .
\end{equation}
In this expression for $P_n$, the index $q$ is the number of connected
subdiagrams that are not cut (this can be seen from the fact that it
comes from the absorptive correction, whose sole role is to preserve
unitarity). 

The term of fixed $p$ and $q$ in this formula can therefore be
interpreted as the probability of producing $n$ particles from $p+q$
connected subdiagrams, $p$ of which are cut and $q$ of which are not
cut. By summing this probability over $n$, one ``integrates out'' some
degrees of freedom in order to keep only the information about the
probability of having $p$ cut subdiagrams and $q$ that are not cut,
regardless of the number of produced particles. This probability
reads~:
\begin{equation}
{\cal R}_{p,q}=\frac{(-1)^q}{p!q!}\Big(\sum_r\frac{b_r}{g^2}\Big)^{p+q}\; .
\label{eq:Rpq}
\end{equation}
Summing this expression over $q$ from $0$ to $\infty$, one finally
obtains the probability of having $p$ cut subdiagrams~:
\begin{equation}
{\cal R}_p=e^{-\sum_r b_r/g^2}\;\frac{1}{p!}\; \Big(\sum_r\frac{b_r}{g^2}\Big)^p\; .
\label{eq:Rp}
\end{equation}
One therefore sees that the number of cut subdiagrams has a Poissonian
distribution, with an average of $\big<n\big>_{\rm cut}=\sum_r
{b_r}/{g^2}$. Eqs.~(\ref{eq:Rpq}) and (\ref{eq:Rp}) are the essence of
the Abramovsky-Gribov-Kancheli cancellations \cite{AGK}. As one can
see from the above derivation, they are simply a consequence of the
factorization of a generic diagram in terms of its connected
subdiagrams. Therefore, we expect them to be much more general than
the context of reggeons field theories in which they have first
been discussed. Another point should also be obvious at this point~:
in order to obtain the eqs.~(\ref{eq:Rpq}) and (\ref{eq:Rp}), one has
``integrated out'' the number $n$ of produced particles. By doing this,
the infinite sequence $b_1,b_2,b_3,\cdots$ has reduced to the single
combination $\sum_r b_r$. This means that a lot of the dynamical
information about the theory under consideration has been lost in this
process, and that there are certain questions that cannot be answered
anymore by the sole knowledge of the ${\cal R}_{p,q}$'s.

\section{Generating function}
\label{sec:gen}
Let us now consider the generating function ${\cal F}(z)$ per se, and
discuss what it would take to calculate it. The following results on
this question have been established in \cite{GelisV}~:
\begin{itemize}
\item[(i)] Diagrammatically, ${\cal F}(z)$ is the sum of all the cut
  vacuum-vacuum diagrams, where each cut propagator is weighted by a
  factor $z$.
\item[(ii)] At {\bf leading order}, the derivative of $\ln {\cal
    F}(z)$, ${\cal F}^\prime(z)/{\cal F}(z)$ can be calculated from a
  pair of solutions $\Phi_\pm(z|x)$ of the {\sl classical equation of
    motion}. For a scalar field theory with a cubic coupling and a
  source $j$, the classical EOM reads
  \begin{equation}
    (\square_x+m^2)\Phi_\pm(z|x)+\frac{g}{2}\Phi_\pm^2(z|x)=j(x)\; .
  \end{equation}
  The expression of ${\cal F}^\prime(z)/{\cal F}(z)$ is simpler if
  written in terms of the Fourier modes $f_\pm^{(+)}(z|x^0,\p)$ and
  $f_\pm^{(-)}(z|x^0,\p)$ of these classical fields,
  \begin{equation}
    \Phi_\pm(z|x)\equiv
    \int\frac{d^3\p}{(2\pi)^3 2E_\p}
    \;
    \Big\{
    f_\pm^{(+)}(z|x^0,\p)\,e^{-ip\cdot x}
    +
    f_\pm^{(-)}(z|x^0,\p)\,e^{+ip\cdot x}
    \Big\}\; ,
    \label{eq:fourier-phi}
  \end{equation}
  and reads
  \begin{equation}
    \left.
    \frac{{\cal F}^\prime(z)}{{\cal F}(z)}
    \right|_{_{LO}}
    =
    \int \frac{d^3\p}{(2\pi)^3 2E_\p}\;
    f_+^{(+)}(z|+\infty,\p)\,f_-^{(-)}(z|+\infty,\p)\; .
    \label{eq:dF1}
  \end{equation}
\item[(iii)] The two solutions $\Phi_\pm$ of the classical EOM must obey the following
  boundary conditions~:
  \begin{eqnarray}    
    &&
    f_+^{(+)}(z|x^0=-\infty,\p)=0\; ,
    \quad
    f_-^{(-)}(z|x^0=-\infty,\p)=0\; ,
    \nonumber\\
    &&
    f_-^{(+)}(z|x^0=+\infty,\p)=
    z\,f_+^{(+)}(z|x^0=+\infty,\p)\; ,
    \nonumber\\
    &&
    f_+^{(-)}(z|x^0=+\infty,\p)=
    z\,f_-^{(-)}(z|x^0=+\infty,\p)\; .
    \label{eq:boundary3}
  \end{eqnarray}
  \item[(iv)] From unitarity, ${\cal F}(1)=\sum_n P_n=1$. This property
  serves as the initial condition for going from ${\cal
  F}^\prime(z)/{\cal F}(z)$ to ${\cal F}(z)$, by writing~:
  \begin{equation}
    {\cal F}(z)=\exp\left\{\int\limits_1^z d\tau\,\frac{{\cal F}^\prime(\tau)}{{\cal F}(\tau)}\right\}\; .
  \end{equation}
\end{itemize}
Unfortunately, finding the pair of solutions of the classical equation
of motion that obey the boundary conditions of
eqs.~(\ref{eq:boundary3}) is a difficult numerical problem, that has
not yet been studied in this context. However, as we shall see in the
next section, one can obtain from here a formula for the average
multiplicity which is much easier to evaluate numerically.

\section{Average multiplicity at leading order}
\label{sec:nbar}
\subsection{General method}
The moments of the distribution of multiplicities, in particular the
average multiplicity $\big<n\big>\equiv\sum_n n P_n$, enjoy a special
status because extra simplifications occur in their calculation. Let
us consider the multiplicity since it is the simplest one. One can get
the multiplicity from the generating function as
\begin{equation}
\big<n\big>={\cal F}^\prime(1)=\frac{{\cal F}^\prime(1)}{{\cal F}(1)}\; .
\end{equation}
(The second equality is of course simply due to ${\cal F}(1)=1$.)
Therefore, calculating $\big<n\big>$ is a special case -- with $z=1$
-- of the calculation of ${\cal F}^\prime(z)/{\cal F}(z)$. At $z=1$,
the third and fourth of the boundary conditions in
eqs.~(\ref{eq:boundary3}) simply reduce to
\begin{eqnarray}
&&
    f_-^{(+)}(1|x^0=+\infty,\p)=
    f_+^{(+)}(1|x^0=+\infty,\p)\; ,
    \nonumber\\
    &&
    f_+^{(-)}(1|x^0=+\infty,\p)=
    f_-^{(-)}(1|x^0=+\infty,\p)\; ,
    \label{eq:boundary4}
\end{eqnarray}
which means that the fields $\Phi_+$ and $\Phi_-$ are equal at large
positive times, as well as their first time derivative. Since the
classical equation of motion is deterministic, this implies that
these two classical fields are equal at all times. The first two
boundary conditions in eqs.~(\ref{eq:boundary3}) then imply that
$\Phi_\pm$ are vanishing at large negative times, as well as their
first time derivative.

Therefore, at leading order, the multiplicity $\big<n\big>$ is given
by eqs.~(\ref{eq:fourier-phi}) and (\ref{eq:dF1}) with $\Phi_+=\Phi_-$
the {\sl retarded solution of the classical EOM with a null initial
condition at $x_0=-\infty$}. One should emphasize the following: it
was obvious from the beginning that the multiplicity at leading order
would be somehow related to solutions of the classical equation of
motion -- since it involves only tree diagrams at this order -- but it
is a non-trivial result that this is with retarded boundary
conditions. The retarded nature of the boundary condition is crucial
in practice, in order to solve this problem numerically.

\subsection{Gluon multiplicity at leading order}
\begin{figure}[htbp]
\begin{center}
\resizebox*{!}{5cm}{\includegraphics{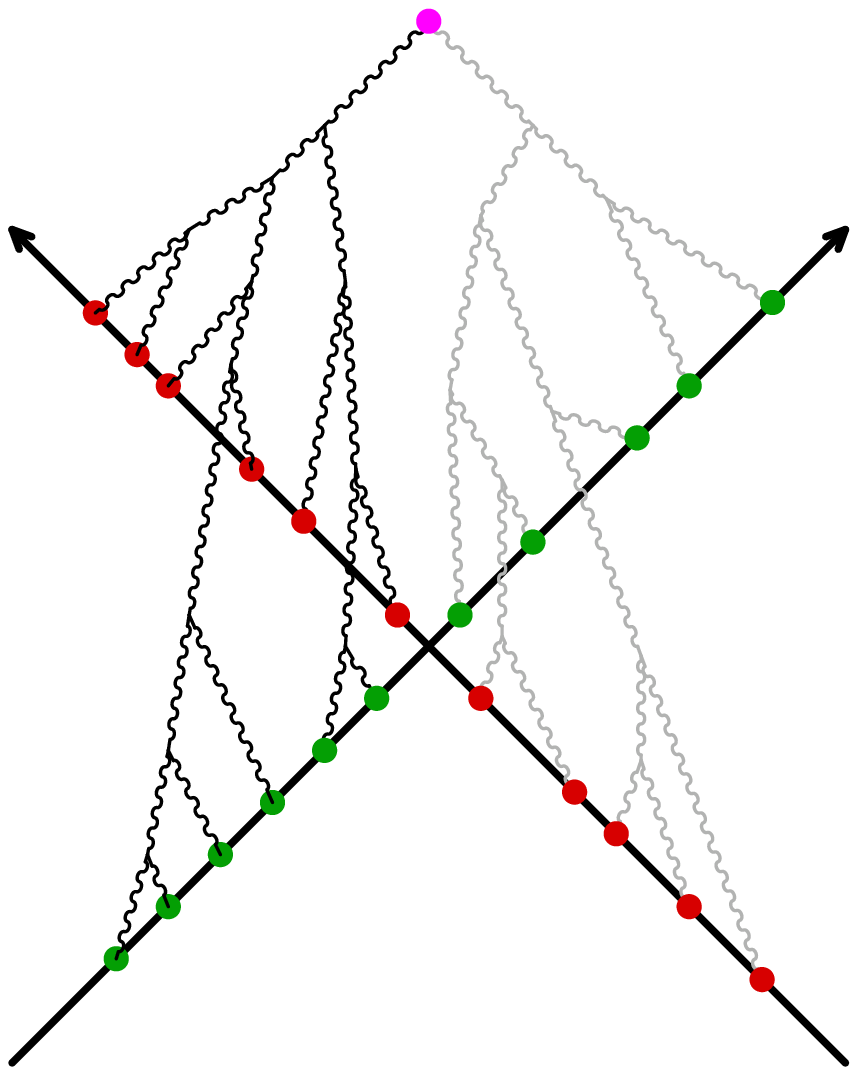}}
\hglue 15mm
\resizebox*{!}{4.5cm}{\includegraphics{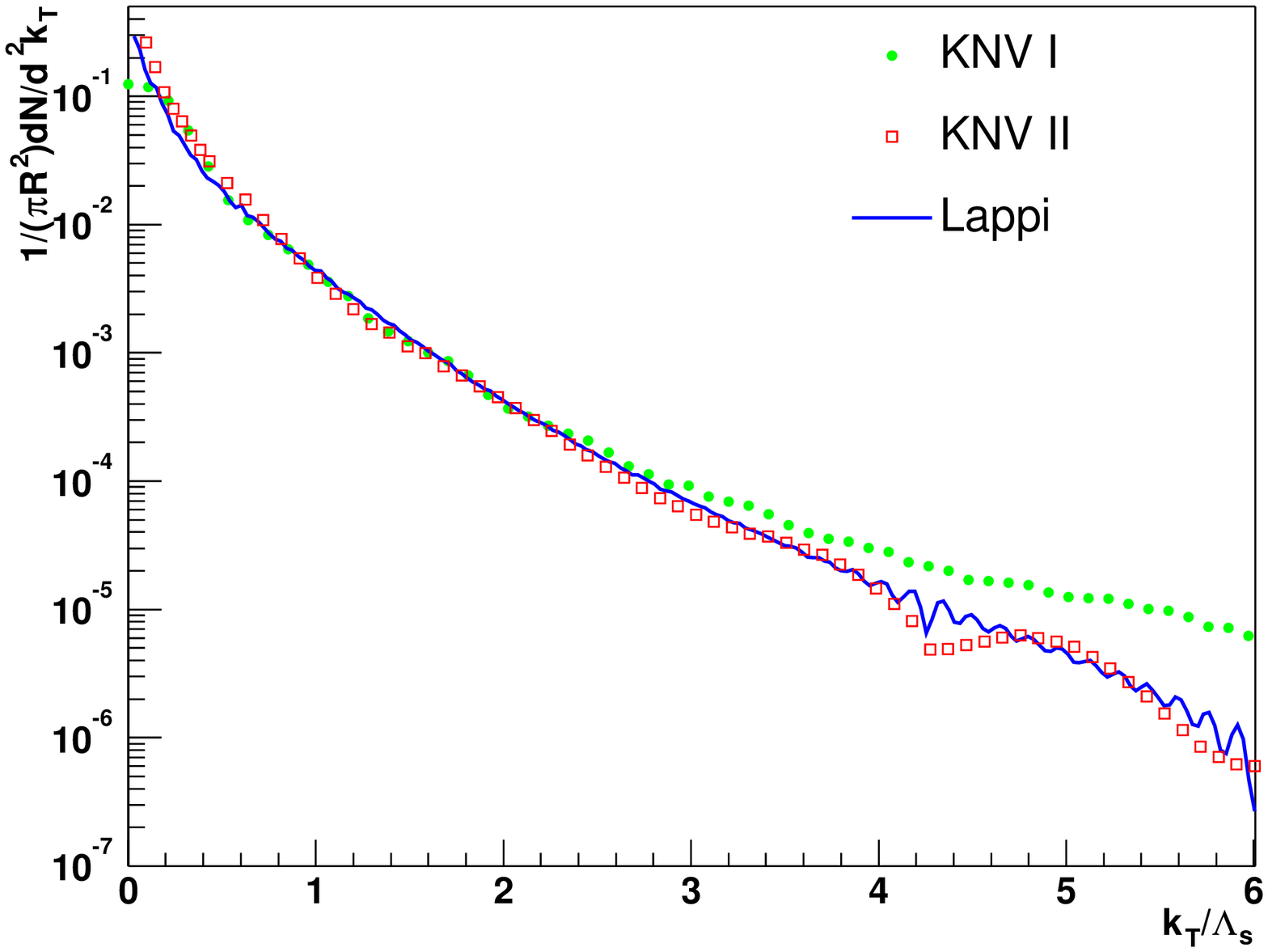}}
\end{center}
\caption{\label{fig:gluons-lo}Left~: space-time representation of the
  diagrams involved in the calculation of the gluon multiplicity at
  leading order via eq.~(\ref{eq:gluon-LO}) (one factor of
  $A_{_R}^\mu$ is represented in black and the other in grey). Right~:
  the resulting gluon spectrum -- the plotted quantity is
  $dN/d^2\k_\perp$.}
\end{figure}
Going back to QCD, it is straightforward to generalize the previous
results to the case of gluon production. The inclusive gluon spectrum
is given at leading order by
\begin{equation}
E_\p
\frac{d \big<n_{\rm gluons}\big>_{_{LO}}}{d^3\vec\p}=
\frac{1}{16\pi^3}\sum_\lambda\int_{x,y}\; { e^{ip\cdot (x-y)}}\;
\square_x\square_y\;
 { \varepsilon_\lambda\cdot A_{_R}(x)\;
\varepsilon_\lambda\cdot A_{_R}(y)}\; .
\label{eq:gluon-LO}
\end{equation}
In this formula, $A^\mu_{_R}(x)$ is the {\sl retarded solution of the
  classical Yang-Mills equations} -- in the presence of the sources
$\rho_{1,2}$ -- and with $A^\mu_{_R}=0$ and $\partial^0 A^\mu_{_R}=0$
at $x_0=-\infty$. This problem was solved numerically in \cite{GLO}.
The diagrams\footnote{The propagators that appear in this diagrammatic
  representation are not Feynman propagators, but retarded
  propagators.} that are resummed by eq.~(\ref{eq:gluon-LO}) are
represented in figure \ref{fig:gluons-lo}, as well as the resulting
gluon spectrum.

\subsection{Quark production at leading order}
Similarly, the production of quarks at leading order has been
considered in \cite{QLO}. Note that we use ``leading order'' somewhat
abusively here, since quark production is strictly speaking a
Next-to-Leading Order effect -- indeed, in the saturated regime, the
number of produced gluons scales like $g^{-2}$ while the number of
produced quarks scales like $g^0$.
\begin{figure}[htbp]
\begin{center}
\resizebox*{!}{5cm}{\includegraphics{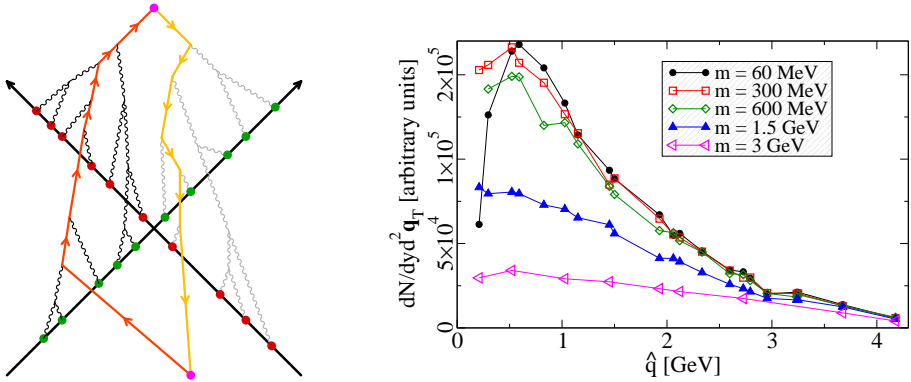}}
\hglue 10mm
\resizebox*{!}{4.5cm}{\includegraphics{AA-pspectqs20p.ps}}
\end{center}
\caption{\label{fig:quarks-lo}Left~: space-time representation of the
diagrams involved in the calculation of the quark multiplicity at
leading order. Right~: the resulting quark spectra for various quark
masses.}
\end{figure}
For quark production at leading order, one must use the following
formula,
\begin{equation}
E_\p
\frac{d{{\big<n_{\rm quarks}\big>_{_{LO}}}}}{d^3\vec\p}=
\frac{1}{16\pi^3}\int_{x,y}\int_\q\; { e^{ip\cdot (x-y)}}\;
(i\slpartial_x-m)(i\slpartial_y+m)\;
{ \overline{\psi_\q}(x)\psi_\q(y)}\; ,
\label{eq:q-lo}
\end{equation}
where $\psi_\q(x)$ is the retarded solution of the Dirac equation --
with the classical field obtained in the calculation of gluon
production in the background -- with a free negative energy spinor,
$v(\q)e^{iq\cdot x}$, as the initial condition. The diagrams involved
in eq.~(\ref{eq:q-lo}) are sketched in the left of figure
\ref{fig:quarks-lo}. Also represented in the right part of this figure
is the resulting quark spectrum, for various quark masses.

\section{Further developments}
\subsection{Gluon multiplicity at NLO}
In fact, the production of quarks is one of the pieces -- the simplest
one -- that contribute to particle production at NLO, i.e. at order
$g^0$. In \cite{GelisV}, we have detailed the principles of a full NLO
calculation of the particle yield, in the case of scalar fields. Of
course, things will be more complicated in QCD with gluons, but one
crucial property of the result will survive~: {\sl the calculation of
particle production at NLO can be done from retarded solutions of the
classical EOM and retarded solutions of the EOM of a small fluctuation
on top of the classical field}. The crucial point again is the fact
that these objects are needed with retarded boundary conditions, which
means that it is a problem which is tractable numerically in a
straightforward way.
\begin{figure}[htbp]
\begin{center}
\resizebox*{!}{5cm}{\includegraphics{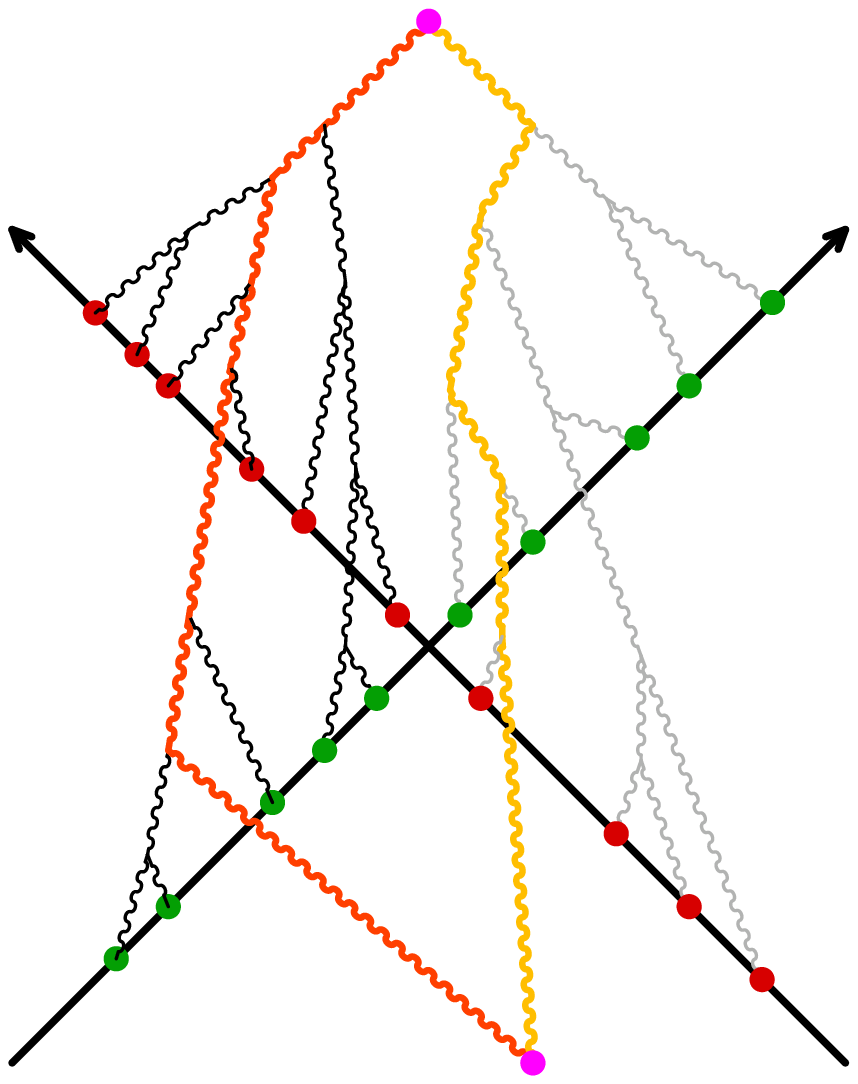}}
\hglue 15mm
\resizebox*{!}{5cm}{\includegraphics{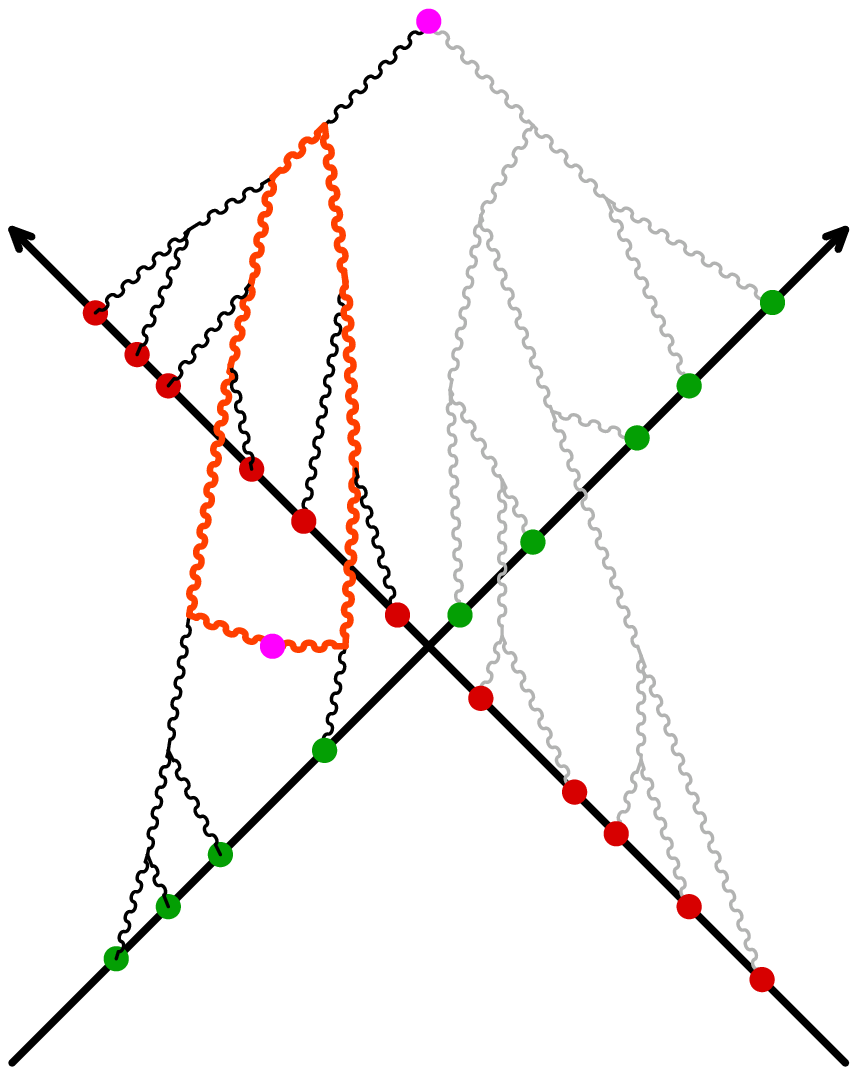}}
\end{center}
\caption{\label{fig:gluons-nlo}Space-time representation of the
diagrams involved in the gluon multiplicity at NLO.}
\end{figure}

Two types of topologies, sketched in figure \ref{fig:gluons-nlo},
contribute to the gluon multiplicity at NLO \cite{GLV}.  One of them
(left diagram) is very similar to that already encountered in quark
production -- it corresponds to the production of pairs of gluons,
and involves retarded solutions for the equation of motion of small
gluonic fluctuations on top of the classical field. The diagram on the
right can be seen as a 1-loop virtual correction to the classical
field -- the field in the complex conjugate amplitude remaining the
tree-level one.  It was shown in \cite{GelisV} that the latter
contribution can also be expressed in terms of retarded solutions of
the small fluctuations equation. Therefore, the result according to
which the inclusive multiplicity can be calculated from retarded
solutions of some equations of motion remain true at NLO. To this
diagram with a virtual gluon loop, one must add two similar diagrams,
respectively with a quark loop and a ghost loop (only in gauges that
have ghosts for the latter).

One additional issue arises when one considers these one-loop
corrections to the gluon yield: some of the contributions have a
divergence of the form $\alpha_s\int dx/x$ which is reminiscent of the
divergences already resummed by the JIMWLK evolution of the
distribution of sources $W[\rho_1]$ and $W[\rho_2]$ for the two
projectiles. For the CGC framework to be self-consistent, one must
prove that these divergences that appear in the gluon yield for fixed
$\rho_1$ and $\rho_2$ can be absorbed in the evolution of the
$W[\rho_{1,2}]$ \cite{GLV}.  

\subsection{Survival probabilities}
One can also consider exclusive processes in this framework. For
instance, instead of the plain -- fully inclusive -- probability $P_n$
of producing $n$ particles, one may define a probability $P_n(\Omega)$
of producing $n$ particles in a certain region $\Omega$ of the
phase-space {\sl and none outside of $\Omega$}, and construct a
generating function for these exclusive probabilities, $ {\cal
F}_\Omega(z)\equiv\sum_{n=0}^\infty P_n(\Omega)\,z^n\; .  $ One can
show \cite{GelisV1} that this generating function can in principle be
calculated at leading order by methods that are similar to those
described in section \ref{sec:gen}, modulo two differences.
\begin{itemize}
\item[(i)] The boundary conditions for the two classical fields in
terms of which on can express ${\cal F}_\Omega^\prime(z)/{\cal
F}_\Omega(z)$ read~:
\begin{eqnarray}    
    &&
    f_+^{(+)}(z|x^0=-\infty,\p)=0\; ,
    \quad
    f_-^{(-)}(z|x^0=-\infty,\p)=0\; ,
    \nonumber\\
    &&
    f_-^{(+)}(z|x^0=+\infty,\p)=
    z\,\Omega(\p)\,f_+^{(+)}(z|x^0=+\infty,\p)\; ,
    \nonumber\\
    &&
    f_+^{(-)}(z|x^0=+\infty,\p)=
    z\,\Omega(\p)\,f_-^{(-)}(z|x^0=+\infty,\p)\; ,
    \label{eq:boundary5}
  \end{eqnarray}
where $\Omega(\p)$ is a function which is 1 in the region $\Omega$ and
zero outside.
\item[(ii)] {\sl The ``integration constant'' ${\cal F}_\Omega(1)$ is no
longer unity}. In fact, this quantity is the total probability of not
producing particles outside of $\Omega$. It will appear as a prefactor
in all the probabilities $P_n(\Omega)$, and it can therefore be
interpreted as a {\sl survival probability} for the empty region
outside $\Omega$. 

\end{itemize}

\section{Conclusions}
Multiparticle production in field theories coupled to external
time-dependent sources -- e.g. in the Color Glass Condensate framework
for hadronic collisions at high energy -- exhibits some non-trivial
features when these sources are as strong as the inverse coupling:
even in the weak coupling regime, one must resum at each order an
infinite set of diagrams. Quite generically, one recovers in this type
of model the combinatoric relations among the probabilities of cut
subdiagrams that lead to the AGK cancellations. 

At leading order, both the generating function of the probability
distribution and the average multiplicity can be calculated from
solutions of the classical equation of motion in the presence of the
external sources. However, while these solutions must be found with
complicated boundary conditions in the case of the generating
function, the problem can be reduced to finding retarded solutions in
the case of the multiplicity, leading to straightforward algorithms
for calculating the multiplicity numerically. This has been done at
leading order in the CGC framework for the production of gluons and
quarks in nucleus-nucleus collisions.

The present study can be extended in several directions. One of these
extensions is the production of gluons at NLO (NLO particle production
has in fact been studied in \cite{GelisV}, but the techniques
developed there must be extended to gauge theories). Another
extension is the study of exclusive reactions, where one enforces some
constraints on the phase-space of the produced particles.


\end{document}